\documentclass[
reprint,
amsmath,amssymb,
aps,
prb
]{revtex4-2}
\DeclareUnicodeCharacter{2061}{}
\DeclareUnicodeCharacter{2212}{}

\usepackage{graphicx}% Include figure files
\usepackage{dcolumn}% Align table columns on decimal point
\usepackage{bm}% bold math
\usepackage[table]{xcolor}

\newcommand{\ti}{\textit}
\newcommand{\tb}{\textbf}

\begin{document}
\title{A Hartree--Fock Analysis of the Finite Jellium Model}

\author{Michael Píro}
\email{michael.piro@fjfi.cvut.cz}
\author{Jaroslav Hamrle} 
\affiliation{
Department of Solid State Engineering, 
Faculty of Nuclear Sciences and Physical Engineering, Czech Technical University, Trojanova 13, Prague, Czech Republic
}

\date{\today}% It is always \today, today,
             %  but any date may be explicitly specified

%Use showkeys class option if keyword

\begin{abstract}
A Hartree–Fock analysis of the ground-state electronic structure of the finite spherical jellium model is carried out for systems containing up to $520$ electrons in a positive background field with densities ranging from $10^{-3}$ to $1$. The study focuses on quantifying the effects of confinement on the local-density models of the exchange and kinetic energies used in orbital-free density-based quantum computation methods. Significant discrepancies are observed between the energies obtained from the Hartree–Fock approximation and those predicted by the local density approximation (LDA) and the Thomas--Fermi model (TF) evaluated at the computed electron densities, both in the inner region and on the surface of the system. To reconcile these differences, refined expressions for the local one-electron energy densities, parametrized by the system's size and background charge density, are proposed. These models are also compared with commonly used gradient-based energy functionals. 
\end{abstract}

\maketitle

\section{Introduction}
Density Functional Theory (DFT) has become the cornerstone of modern first-principles quantum‑mechanical simulations of the electronic structure in atoms, molecules and solids. Through the foundational theorems of Hohenberg and Kohn, the ground‑state energy of an interacting many‑electron system can in principle be expressed as a functional of the electron density alone, circumventing the need to solve the full many‑body Schrödinger equation. In practical implementations, the Local Density Approximation (LDA) — which assumes that the exchange energy density is the same as that of an infinite homogeneous electron gas — remains one of the most widely used approximations.
\par
However, the assumption of uniform distribution of the electron density breaks down in finite systems, where the density exhibits strong spatial oscillations and rapid decay near the surface, rendering the simple uniform-density model insufficient \cite{PhysRevLett.57.862, cohen2012challenges}. To mitigate these deficiencies, a variety of more sophisticated approximations have been developed. The most direct extension is the Generalized Gradient Approximation (GGA), such as the Perdew--Burke--Ernzerhof (PBE) functional, which augments the local density with dependence on its gradient \cite{perdew1996generalized}. Beyond GGA, more advanced functionals that depend on higher-order derivatives or kinetic energy density, as well as hybrid functionals that mix in a fraction of exact exchange from wavefunction-based methods, have been introduced to further refine the treatment of exchange.
\par
At the same time, for very large systems, conventional orbital-based DFT becomes computationally expensive, and methods such as Orbital‑Free Density Functional Theory (OF‑DFT) \cite{mi2023orbital}, Quantum Hydrodynamics (QHD) \cite{PhysRevB.93.205405, PhysRevB.107.205413}, or even Quantum-Corrected classical models (QCM) \cite{esteban2012bridging, zapata2015quantum} have to be employed. Since these approaches rely strictly on the electron density, the kinetic energy contribution of the hamiltonian must also be provided as an explicit density functional.
\par
We revisit these fundamental issues by performing a Hartree–Fock analysis (HF) of the ground-state electronic structure of the finite spherical jellium model and studying the effect of its confinement on the local exchange and kinetic energies functionals compared to their infinite jellium analogs. This model has been particularly successful in describing the delocalized conduction electrons of alkali-metal clusters, where the discrete ion background is approximated as a uniform positive charge distribution \cite{PhysRevB.1.4555, brack1993physics}. These previous studies addressed electronic spectra primarily by DFT methods \cite{PhysRevB.29.1558, PhysRevLett.52.1925, PhysRevB.32.5023, knight1987alkali}, with Hartree–Fock calculations available only for a few closed-shell sodium clusters \cite{PhysRevB.45.11283, hansen1993exchange, madjet1995comparative}. In contrast, the present work aims to develop general analytic formulas for the effect of confinement by exploring systems with up to $520$ electrons and background densities ranging from $10^{-3}$ to~$1$.
\par
In order to accurately build the model from the ground up, we first study the behavior of energies and wave functions of both the one-electron and many-electron systems in the jellium potential generated by the background charge. Only then do we proceed to construct the local models of the one-electron energy densities parameterized by the system's size and background charge.

\section{Model description}
In the finite spherical jellium model, the system is described as a solid ball of radius $R$ composed of $N$ atoms. Each atom contributes a positively charged ionic core (consisting of the nucleus and tightly bound inner electrons) and $\nu$ delocalized electrons. The positive charge is assumed to be uniformly distributed over the ball, resulting in the ionic charge density:
\begin{align}
    n_{\text{I}}(\tb{r}) = 
    \begin{cases}
        n_{\text{I}} & |\tb{r}| \leq R \\
        0 & |\tb{r}| > R,
    \end{cases}
\end{align}
where $n_{\text{I}}$ is the background density constant. The total positive charge, given by $Q = \nu N$, can then be expressed as $Q = \frac{4}{3}\pi R^3n_{\text{I}}$. Taking into account the original crystalline arrangement of the atoms, the radius $R$ is related to the number of atoms by $R = r_sN^{1/3}$, where $r_s$ is the Wigner-Seitz radius. Substituting this into the expression for $Q$, the ionic density constant becomes:
\begin{align}
    n_{\text{I}} = \frac{3}{4\pi}\frac{\nu}{ {r_s}^3}. \label{eq:positive_background_const}
\end{align}
As shown in the Supplementary Material, Section S1, this positive background creates the electrostatic \ti{jellium potential} $V(r)$:
\begin{align}
    V(r) = \begin{cases}
        \frac{Q}{2R^3}(r^2 - 3R^2) & r \leq R \\
        -\frac{Q}{r} & r > R.
    \end{cases} \label{eq:potential_R}
\end{align}
Its shape is depicted in Fig. \ref{fig:one_el_wf}. In this study, we analyze systems with densities $n_\text{I} \in [10^{-3}, 1]$ and $\nu = 1$. This corresponds to $r_s \in [0.62, 6.20]$, covering a wide range of different materials.
\par
The system of a single electron in the jellium potential is spherically symmetric. The electron wave function is therefore separated into its angular and radial components:
\begin{align}
    \psi_{nlm}(r,\theta,\phi) = \frac{1}{r}u_{nl}(r)Y_l^m(\theta,\phi), \label{eq:decomposition}
\end{align}
where $n$, $l$, and $m$ denote the principal, orbital, and magnetic quantum numbers, respectively. Since the angular wave functions $Y_l^m(\theta,\phi)$  are determined analytically and are independent of the shape of the radial potential $V(r)$, we further examine only the radial components.
\par
For the many-electron problem, the electron-electron interaction is added via the potential $V_{\text{e-e}}$ accounting for all pairwise interactions:
\begin{align}
    V_{\text{e-e}}(\tb{r}_1,\dots,\tb{r}_N) = \sum_{\begin{subarray}{l}i,j = 1 \\
    \ i < j\end{subarray}}^NV_2(\tb{r}_i,\tb{r}_j) = \sum_{\begin{subarray}{l}i,j = 1 \\
    \ i < j\end{subarray}}^N\frac{1}{|\tb{r}_i-\tb{r}_j|}.
\end{align}
Next, the spatial one-electron wave function is extended to include spin. We consider only spin-symmetrical solutions. The total many-electron wave function is given by the Slater determinant. The interaction potential then decomposes into two distinct contributions: the Hartree potential $\hat{V}_{\text{H}}$ and the exchange potential $\hat{V}_\text{x}$, defined as:
\begin{gather}
    \hat{V}_{\text{H}}\psi_i(\tb{r}) = \sum_{\begin{subarray}{l}j = 1 \\
    j\neq i\end{subarray}}^N\int_{\mathbb{R}^{3}}\psi_j^*(\tb{r}')V_2(\tb{r},\tb{r}')\psi_j(\tb{r}')d\tb{r}'\psi_i(\tb{r}), \label{eq:Hartree_potential_op} \\
    \hat{V}_\text{x}\,\psi_i(\tb{r}) = -\sum_{\begin{subarray}{l}j = 1 \\
    j\neq i\end{subarray}}^N\delta_{s_i s_j}\int_{\mathbb{R}^{3}}\psi_j^*(\tb{r}')V_2(\tb{r},\tb{r}')\psi_i(\tb{r}')d\tb{r}'\,\psi_j(\tb{r}), \label{eq:exchange_potential_op}
\end{gather}
where $s_i$ denotes the spin of the $i$-th electron. It is important to note that the inclusion of self-interaction terms (contributions with $i=j$) in both $\hat{V}_{\text{H}}$ and $\hat{V}_\text{x}$ results in their mutual cancelation, leaving the total electron–electron interaction potential unchanged. This small change is useful for the simplification of further calculations.
\par
The one-electron wave functions are again decomposed into their radial and angular parts, as in Eq.\ (\ref{eq:decomposition}), now with the addition of the spin quantum number $s$ to the radial functions. To preserve spherical symmetry of the solution, electrons in partially filled shells are averaged over all quantum numbers $m$. In this setting, the Hartree and exchange potentials can be expressed as operators acting solely on the radial wave functions $u_{nl}^s$ (see Supplementary Material, Section S2):
\begin{align}
    \hat{V}_{\text{H}} u^s_{nl}(r) =& \nonumber\\
    = \sum_{(n',\,l',\,s')}& N^{s'}_{n'l'} \int_0^{+\infty}u^{s'}_{n'l'}(r')^2\frac{1}{\max(r, r')}dr'u^s_{nl}(r) \label{eq:Hartree_potential_def}
\end{align}
and:
\begin{align}
    \hat{V}_\text{x}&u^s_{nl}(r) = -\sum_{(n',\,l')}N^s_{n'l'}\sum_{L=|l-l'|}^{l + l'}\begin{pmatrix}
    l' & L & l \\
    0 & 0 & 0
    \end{pmatrix}^2 \cdot \nonumber\\
    \cdot& \int_0^{+\infty}u^s_{n'l'}(r')u^s_{nl}(r')\frac{\min(r, r')^L}{\max(r, r')^{L+1}}dr'u^s_{n'l'}(r), \label{eq:radial_exchange_op_def}
\end{align}
where $N^s_{nl}$ is the number of electrons in the shell $(n, l)$ with spin $s$.

\section{One-Electron System}
\begin{figure}[t]
\includegraphics[width=\columnwidth]{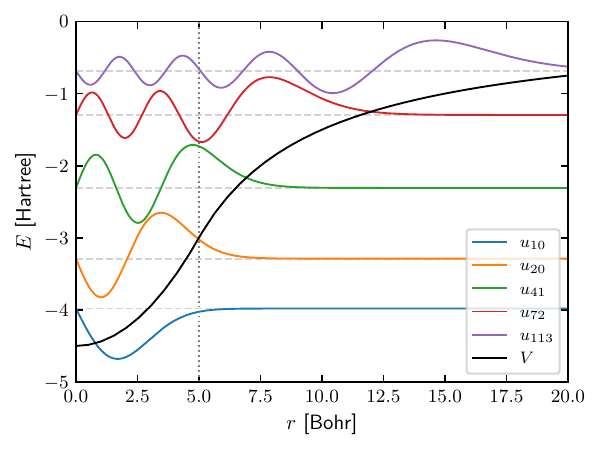}
\caption{\label{fig:one_el_wf} Jellium potential $V$ and five representative one-electron radial wave functions $u_{nl}$ for the system of radius $R = 5$ and background charge $Q = 15$. Horizontal dashed lines indicate corresponding energy levels. Vertical dotted line marks the system radius $R$.}
\end{figure}

In order to gain basic insight into the energy spectrum and properties of the wave functions, we first focus on solving the problem of a single electron in the jellium potential $V$. This system can be viewed as a large atom with a nucleus radius $R$ and a nuclear charge $Q$. As a representative example, we consider a system with $R = 5$ and $Q = 15$. Energies are computed for different values of the orbital quantum number $l$. The principal quantum number $n$ is then assigned according to the hydrogenic convention, where for each $n$, the allowed values of $l$ range from $0$ to $n - 1$. The shapes of several radial wave functions $u_{nl}$ are shown in Fig. \ref{fig:one_el_wf}.
\par
\begin{figure*}
\includegraphics[width=\textwidth]{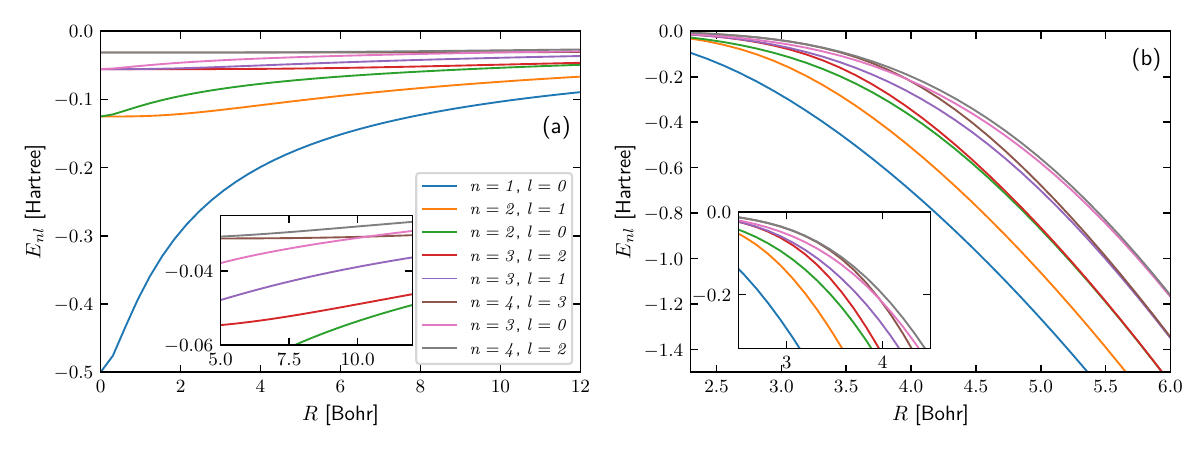}
\caption{\label{fig:energy_dependencies}Dependence of the one-electron energy levels on the radius of the nanoparticle. (a) Positive background charge is fixed at $Q = 1$. (b) Positive background charge density is fixed at $n_\text{I} = 10^{-2}$.}
\end{figure*}

At low energies, electrons are predominantly localized within the radius $R$. However, at higher excited states the wave functions extend significantly beyond the boundary with a gradual increase in both the amplitudes and the wavelengths. The effect of the finite nucleus size $R$ on the energy spectrum is depicted in Fig. \ref{fig:energy_dependencies}.
We analyze two different settings: First, the positive background charge is fixed at $Q = 1$. Second, the positive charge density is fixed at $n_\text{I} = 10^{-2}$. In the first case, we start at $R = 0$ where the spectrum corresponds to that of the hydrogen atom. As the radius increases, the degeneracy in $l$ is lifted, breaking the spherical symmetry. In contrast, certain levels, such as $E_{20}$ and $E_{32}$, or $E_{30}$ and $E_{42}$, tend to converge.
This points to a different degeneracy pattern, namely the one of the harmonic oscillator. This effect is even more profound in the second studied case with a constant positive charge density. Here, the lower-laying electrons converge to an equidistant energy spectrum for increasing radius $R$.
\par
Based on these findings and on the shapes of the wave functions, we distinguish two electron energy classes: The energy levels corresponding to the electrons located mainly within the radius of the nanoparticle $R$, which we call \textit{harmonic-like}, and energies for which the electrons have a significant part of the wave function outside the ionic core. These are called \textit{hydrogen-like}.

\section{Many-electron system}
\subsection{Energy level ordering}
The first aspect of the many-electron system that we examine is the energy level ordering. By incrementally solving the Hartree–Fock equations for increasing electron number $N$, we obtain sequences of occupied orbitals. These vary slightly for different background densities $n_\text{I}$, but exhibit a common pattern: For every principal quantum number $n$, the orbital quantum number $l$ is filled in descending order from its maximum allowed value. This reflects the energetic preference for orbitals with the smallest number of radial nodes. Equivalently, the system maximizes the angular kinetic energy while minimizing the radial kinetic energy. For example, the ordering sequence for $n_\text{I} = 10^{-2}$ starts as follows:
\begin{align}
    1s\ 2p\ 2s\ 3d\ 4f\ 3p\ 5g\ 3s\ 4d\ 6h\ 5f\ 7i\ 4p\ \dots
\end{align}
\par
\begin{figure}[h]
\includegraphics[width=\columnwidth]{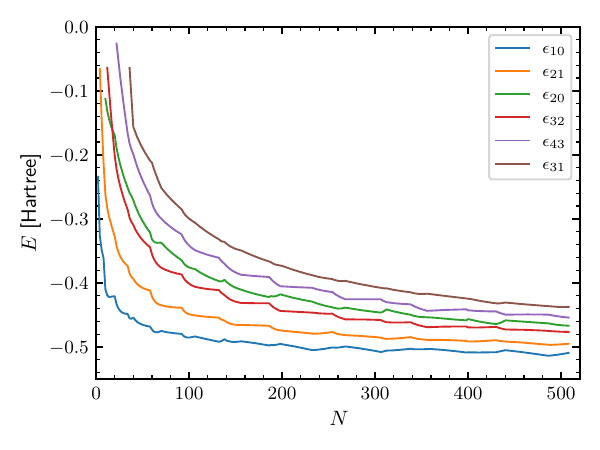}
\caption{\label{fig:energy_levels}Evolution of the one-electron Hartree energies $\epsilon_{nl}$ as a function of electron number $N$ for the six lowest-energy orbital configurations, $n_\text{I} = 10^{-2}$.}
\end{figure}

The evolution of the one-electron Hartree energies $\epsilon_{nl}$ with $N$, shown in Fig.~\ref{fig:energy_levels} for $n_\text{I} = 10^{-2}$, further illustrates the structure of the energy level sequence.
A notable feature is the reordering of the $s$ and $d$ orbitals. We see that the $3d$ orbital rapidly drops below the $2s$ level, despite the latter being filled earlier. This inversion can also be observed between levels $3s$ and $4d$ and is analogous to the well-known behavior in atomic systems, where the $3d$ orbitals become energetically favorable relative to $4s$.

\subsection{Electron Density}
\begin{figure*}
\includegraphics[width=\textwidth]{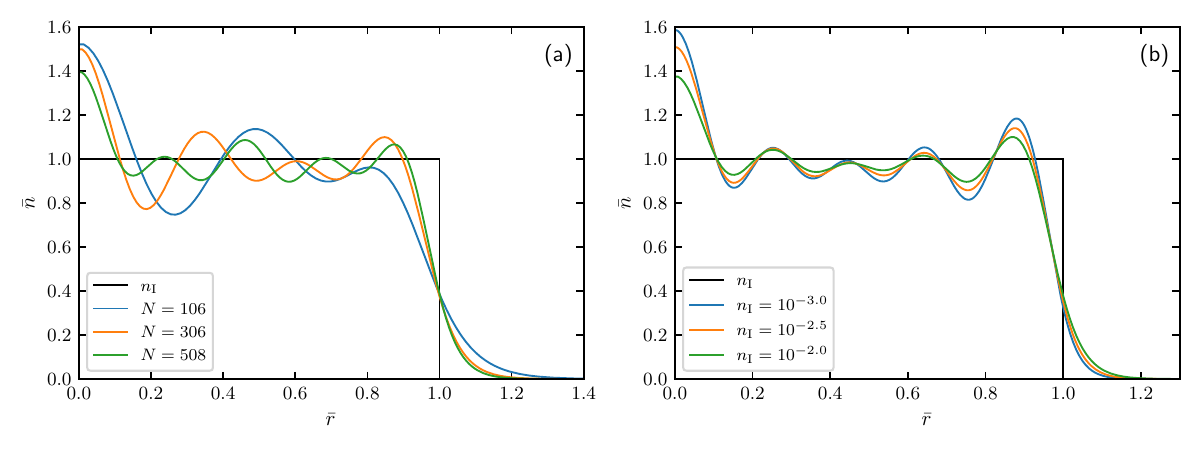}
\caption{\label{fig:densities}Radial electron density distributions $n(r)$, normalized by the background charge density $n_{\text{I}}$, as functions of the normalized radius $\bar{r} = r/R$. (a) Background charge density $n_{\text{I}} = 10^{-2}$, $N$ corresponds to three different systems with fully filled orbitals. (b) Comparison of normalized radial electron densities of systems with fully filled orbitals with $N = 486$ electrons, evaluated at different background charge densities.}
\end{figure*}

Next, we study the spatial distribution of the electron density. Due to the spherical symmetry of the model, the total electron density depends only on the radial coordinate and is given by:
\begin{align}
    n(r) = \sum_{(n,\,l,\,s)}N^s_{nl}\frac{u^s_{nl}(r)^2}{4\pi r^2}.\label{eq:electron_density_def}
\end{align}
Fig. \ref{fig:densities}(a) shows the normalized electron density profiles, defined as $\bar{n}(r) = n(r)/n_{\text{I}}$, as functions of the normalized radius $\bar{r} = r/R$, for $n_\text{I} = 10^{-2}$ and three different numbers of electrons $N$.
All profiles exhibit pronounced oscillations around the positive background density. The number of nodes increases approximately linearly with radius $R$. The asymptotic behavior of the position of the last local maximum relative to $R$, characterizing the beginning of the effective electron gas surface, can be estimated by the following relation:
\begin{align}
    \bar{r}_\text{lm} &= 1 - 0.53\exp(-0.19N^{1/3}). \label{eq:r_lm_approx}
\end{align}
Comparing systems with the same number of electrons but different background densities (see Fig. \ref{fig:densities}(b)) reveals that the oscillations of $\bar{n}$ are stronger for lower densities $n_\text{I}$.

\subsection{Energy Contributions}
Another important feature of the system we analyze is the total energy and its individual components. Based on their scaling behavior and relative magnitude, the energy contributions can be grouped into two categories: dominant and subdominant terms. The dominant components include the energy of the interaction between the electrons and the positively charged ionic background $E_{\text{e-I}}$, the Hartree part of the electron–electron interaction $E_{\text{H}}$, and the energy of the ionic background $E_{\text{I}}$ itself. All these contributions follow an $N^{5/3}$ scaling with the amplitudes depending on the background density as:
\begin{align}
    A_\text{e-I} &\approx -1.886\ n_\text{I}^{1/3}, \\
    A_\text{H} &\approx 0.921\ n_\text{I}^{1/3}, \\
    A_\text{I} &= \frac{(36\pi)^{1/3}}{5}n_\text{I}^{1/3} \approx 0.967\ n_\text{I}^{1/3}.
\end{align}
\par
The subdominant contributions include the total kinetic energy $E_\text{kin}$ and the exchange component of the electron-electron interaction energy $E_\text{x}$. These follow a linear scaling in $N$ with amplitudes depending on the background density as:
\begin{align}
    A_\text{kin} &\approx 2.273\ n_\text{I}^{2/3}, \label{eq:A_kin} \\
    A_\text{x} &\approx -0.658\ n_\text{I}^{1/3}. \label{eq:A_x}
\end{align}
As shown in the previous subsection, the electron density $n$ is not uniformly distributed in finite systems, but oscillates around the background density. However, if we make a crude approximation by setting $n(\tb{r}) = n_\text{I}$, we can use the relations (\ref{eq:A_kin}), (\ref{eq:A_x}) to estimate the one-electron kinetic and exchange energy densities as functions of $n$:
\begin{align}
    \varepsilon_\text{kin}(\tb{r}) &\approx 2.273\ n^{2/3}(\tb{r}), \label{eq:crude_kin} \\
    \varepsilon_\text{x}(\tb{r}) &\approx -0.658\ n^{1/3}(\tb{r}). \label{eq:crude_exchange}
\end{align}
Comparing this to the infinite-system analogs given by the LDA and Thomas–Fermi models as:
\begin{align}
    \varepsilon^\infty_\text{kin}(\tb{r}) &\approx 2.871\ n^{2/3}(\tb{r}), \\
    \varepsilon^\infty_\text{x}(\tb{r}) &\approx -0.739\ n^{1/3}(\tb{r}),
\end{align}
one might conclude that the confinement of the system causes a decrease in the absolute value of both the exchange and kinetic energy densities. Nevertheless, a more detailed analysis presented in the next section shows that the oscillatory structure inside the jellium radius $R$, together with the non-negligible electron spill-out, actually magnifies the effect of these energy-density functions.

\section{Exchange and kinetic energies as functionals of the electron density}
In direct analogy with how LDA is constructed, namely by taking the exchange energy density and the electron density of the infinite homogeneous electron gas evaluated in the Hartree–Fock approximation, we determine how confinement modifies the local exchange density. Similarly, the local kinetic energy density is computed directly from the finite system, allowing us to quantify deviations from the infinite-gas limit.

\subsection{Exchange energy density}
\begin{figure}[h]
\includegraphics[width=\columnwidth]{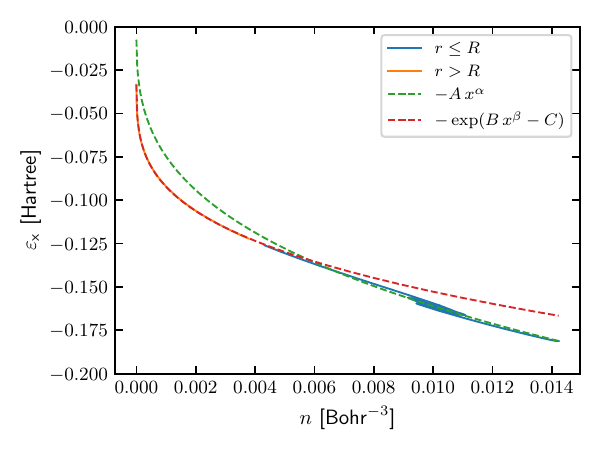}
\caption{\label{fig:eps_x_fit} Dependence of the one-electron exchange energy density  $\varepsilon_\text{x}(r)$ on the electron density $n(r)$ for $n_\text{I} = 10^{-2}$ together with the functions $f$ and $g$ given by Eqs. (\ref{eq:f}), (\ref{eq:g}) for parameters $A = 0.748$, $B = 4.10$, $C = 4.60$, and $\beta = 8.88\times10^{-2}$.}
\end{figure}

The one-electron exchange energy density $\varepsilon_\text{x}(r)$ is defined via the total exchange energy:
\begin{align}
    E_\text{x} = 4\pi\int_0^{+\infty}\varepsilon_\text{x}(r)n(r)r^2dr.
\end{align}
We can compare this with the exchange energy calculated from the Hartree--Fock equations using the defining relation of the exchange potential Eq.\ (\ref{eq:radial_exchange_op_def}). As shown in the Supplementary Material, Section S3, this yields an expression for $\varepsilon_\text{x}$ in terms of the radial wave functions $u^s_{nl}$:
\begin{align}
    \varepsilon_\text{x}(r) &= -\frac{1}{2}\Big[\sum_{(n,\,l,\,s)}N^s_{nl}\sum_{(n',\,l')}N^s_{n'l'}\sum_{L=|l-l'|}^{l + l'}\begin{pmatrix}
    l' & L & l \\
    0 & 0 & 0
    \end{pmatrix}^2 \cdot \nonumber\\
    &\cdot \int_0^{+\infty}u^s_{nl}(r')u^s_{n'l'}(r')\frac{\min(r, r')^L}{\max(r, r')^{L+1}}dr'\cdot \nonumber\\
    &\cdot u^s_{nl}(r)u^s_{n'l'}(r)\Big] \Big/ \sum_{(n,\,l,\,s)}N^s_{nl}u_{nl}^s(r)^2.
\end{align}
To find an approximate form of the exchange energy density as an explicit function of the electron density, we construct a parametric plot of $\varepsilon_\text{x}(r)$ versus $n(r)$ (Fig. \ref{fig:eps_x_fit}).
In the inner region of the system ($r\leq R$), both $n$ and $\varepsilon_\text{x}$ exhibit oscillatory behavior, resulting in a cyclic trajectory in the parametric plot. However, these spirals are sufficiently flat to be well approximated by a monotonic functional dependence. Using the estimate provided by Eq. (\ref{eq:crude_exchange}), we fit the following function:
\begin{align}
    f(x; A) = -A\,x^{1/3}. \label{eq:f}
\end{align}
This provides a good model for most of the inner section, but it loses accuracy near the boundary $r = R$. In the outer region, no valid power law fit that provides a meaningful approximation can be found. To address this, we introduce a second family of exponential-type functions of the form:
\begin{align}
    g(x; B, C, \beta) = -\exp(B\,x^\beta - C). \label{eq:g}
\end{align}
This yields an excellent fit in the outer and near-boundary inner regions, precisely where the power-law approximation fails.
\par
In order to obtain a general formula, we fit both functional classes to the exchange potentials calculated for systems with fully filled electron shells up to $N =~520$ electrons and positive background densities $n_\text{I} \in [10^{-3}, 1]$. It turns out that it is possible to fix one of the coefficients $B$, $C$ for all cases and have the power $\beta$ be only a function of $n_\text{I}$. Setting $C = 4.6$, we found that the parameter $\beta$ is inversely proportional to the positive density and can be well expressed using the following relation:
\begin{align}
    \beta(n_\text{I}) = 1.91\times10^{-5}\ n_\text{I}^{-0.957} + 8.72\times10^{-2}.
\end{align}
\par
\begin{figure*}
\includegraphics[width=\textwidth]{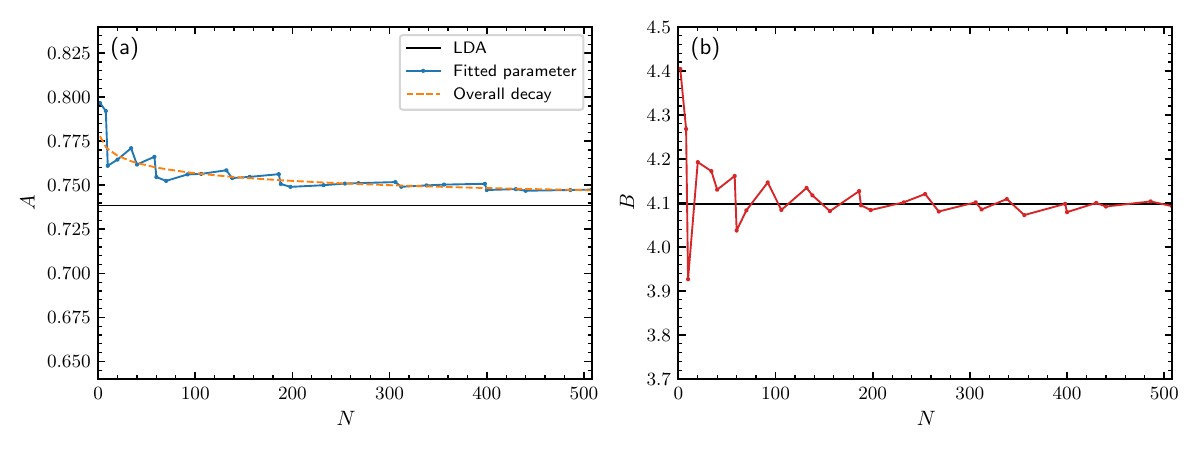}
\caption{\label{fig:params_x} Dependence of the fitted parameters $A$ and $B$ of function $f$ and $g$ given by Eqs. (\ref{eq:f}), (\ref{eq:g}) on the electron number $N$; $n_\text{I} = 10^{-2}$.}
\end{figure*}

Fig. \ref{fig:params_x} shows the remaining coefficients $A$ and $B$ for $n_\text{I} = 10^{-2}$. By comparing the result with the LDA value:
\begin{align}
    A_\text{LDA} = \frac{3}{4}\bigg(\frac{3}{\pi}\bigg)^{1/3} \approx 0.739,
\end{align} we find that the actual dependence on the electron density is stronger for confined systems, but it decays to LDA as $N$ increases. The decay becomes more pronounced for larger background densities, as can be seen from the general estimate for $A$:
\begin{align}
    A(n_\text{I}, N) &= A_1(n_\text{I})\exp(-A_2(n_\text{I})N^{1/3}) + A_\text{LDA}, \\
    A_1(n_\text{I}) &= -5.20\times10^{-3}\log(n_\text{I}) + 4.20\times10^{-2}, \\
    A_2(n_\text{I}) &= 5.58\times10^{-2}\log(n_\text{I}) + 0.339.
\end{align}
The parameter $B$ exhibits oscillatory behavior, but converges toward a well-defined limiting value. The convergence is slightly faster for lower $n_\text{I}$, yet in all cases a deviation below $2\%$ is achieved for $N > 80$. The particular dependence of the limiting value of $B$ on the background charge density is given by:
\begin{align}
    B(n_\text{I}) = 0.262\log(n_\text{I}) + 4.61.
\end{align}
\par
\begin{figure}[h]
\includegraphics[width=\columnwidth]{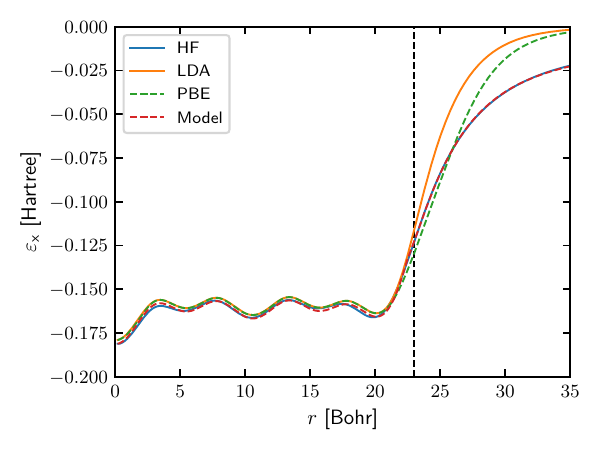}
\caption{\label{fig:eps_x_comparison} Comparison of the local model of the one-electron exchange energy density $\varepsilon_\text{x}$ for parameters $A = 0.748$, $B =~4.10$, and $\beta = 0.089$ with LDA and PBE for $n_\text{I} = 10^{-2}$ and $N = 508$. Vertical dashed line represents the boundary $R$.}
\end{figure}

Finally, for $n_\text{I} \in [10^{-3}, 10^{-3/2}]$ the fitted functions $f$ and $g$ conveniently intersect at the edge of their respective validity ranges. This allows us to define our local model for $\varepsilon_\text{x}$ as the minimum of these two functions:
\begin{align}
    \varepsilon_\text{x}(n) &= \nonumber\\
    =&\min\!\Big(\!-\!A(n_\text{I}, N)n^{1/3},\,-\exp\!\big(B(n_\text{I})\,n^{\beta(n_\text{I})} - 4.6\big)\!\Big).
\end{align}
A comparison of this model with the LDA and PBE functions evaluated at the computed electron densities $n$ for $n_\text{I} = 10^{-2}$ and $N = 508$ is shown in Fig. \ref{fig:eps_x_comparison}.
For higher $n_\text{I}$, the functions $f$ and $g$ no longer intersect, but approach each other in the outer region. Nonetheless, the exponential fit still provides a reasonable improvement for $r > R$. At $n_\text{I} = 1$ and $N > 100$, LDA becomes accurate in both the inner and outer parts, giving a reliable description of the exchange across the entire system.

\subsection{Kinetic energy density}
\begin{figure}[h]
\includegraphics[width=\columnwidth]{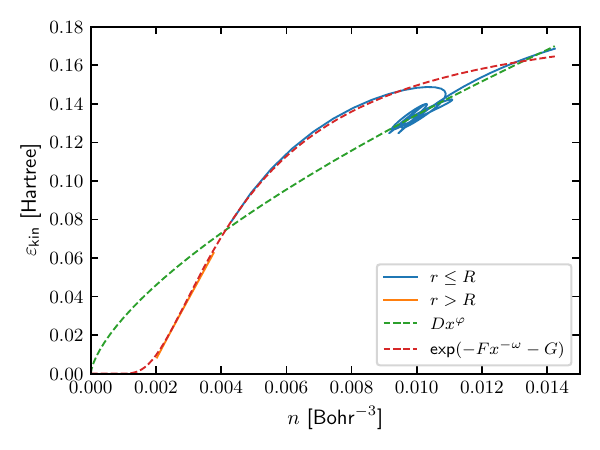}
\caption{\label{fig:eps_kin_fit} Dependence of the one-electron kinetic energy density  $\varepsilon_\text{kin}(r)$ on the electron density $n(r)$ for $n_\text{I} = 10^{-2}$ together with the functions $h$ and $k$ given by Eqs. (\ref{eq:h}), (\ref{eq:k}) for parameters $D = 2.89$, $F = 1.28\times10^{-4}$, $G = 1.68$, and $\omega = 1.60$.}
\end{figure}

Similarly to the exchange energy, we define the one-electron kinetic energy density from the total kinetic energy:
\begin{align}
    E_\text{kin} = 4\pi\int_0^{+\infty}\varepsilon_\text{kin}(r)n(r)r^2dr.
\end{align}
By matching this to the expression for the kinetic energy given in the Hartree--Fock approximation, we arrive at a defining relation for $\varepsilon_\text{kin}$ in terms of the radial wave functions $u_{nl}^s$ (see Supplementary Material, Section S3):
\begin{align}
    \varepsilon_\text{kin}(r) =& -\frac{1}{2}\Big[\sum_{(n,\,l,\,s)}N^s_{nl}\Big(u_{nl}^s(r)\frac{\partial^2u_{nl}^s(r)}{\partial r^2 }\ - \nonumber \\
    &-\frac{l(l+1)}{r^2}u_{nl}^s(r)^2\Big)\Big]\Big/\sum_{(n,\,l,\,s)}N^s_{nl}u_{nl}^s(r)^2.
\end{align}
We now proceed in the same way as for the exchange energy. Fig. \ref{fig:eps_kin_fit} shows the kinetic density plotted against the electron density for $n_\text{I} = 10^{-2}$.
The power-law function of the form:
\begin{align}
    h(x; D) = Dx^{2/3}, \label{eq:h}
\end{align}
motivated by the estimate given in Eq. (\ref{eq:crude_kin}), is only valid for a part of the oscillating region inside $R$. On the other hand, an exponential-type function of the form:
\begin{align}
    k(x; F, G) = \text{exp}(-Fx^{-\omega} - G) \label{eq:k}
\end{align}
provides a good fit from the outer region all the way to the last local maximum of the electron density. We therefore define the local model as a combination of $h$ and $k$ with a smooth transition around this point:
\begin{align}
    \varepsilon_\text{kin} = (1-w)h + w k, \label{eq:eps_kin_model}
\end{align}
where $w$ is the first-order smoothstep function centered around the last peak of the electron density given by Eq.~(\ref{eq:r_lm_approx}) interpolating functions $h$ and $j$ in the symmetric interval $[(3\bar{r}_\text{lm} - 1)R/2, (\bar{r}_\text{lm} + 1)R/2]$.
\par
Again, to obtain the general formula, we fit both functional classes to the exchange potentials calculated for systems with fully filled electron shells up to $N =~520$ electrons and positive background densities $n_\text{I} \in [10^{-3}, 1]$. The parameters $\omega$ and $G$ can be fixed for all $N$, only depending on $n_\text{I}$:
\begin{align}
    G &= -1.56\log(n_\text{I}) - 1.44, \\
    \omega &= 0.241\, n_\text{I}^{-0.292} + 0.693.
\end{align} 
The coefficient $D$ oscillates for lower $N$, but tends to converge to the Thomas--Fermi limit:
\begin{align}
    D_\text{TF} = \frac{3}{10}(3\pi^2)^{2/3}\approx2.87.
\end{align}
Regarding the coefficient $F$, it can be approximated using the following exponential decay:
\begin{align}
    F(n_\text{I}, N) &= -F_1(n_\text{I})\Big(\exp(-F_2(n_\text{I})\,N^{1/3}) + 1\Big), \\
    F_1(n_\text{I}) &= \exp(-1.83\,n_\text{I}^{-0.343}),\\
    F_2(n_\text{I}) &= 0.06\log(n_\text{I}) + 0.67.
\end{align}
\par
\begin{figure}[h]
\includegraphics[width=\columnwidth]{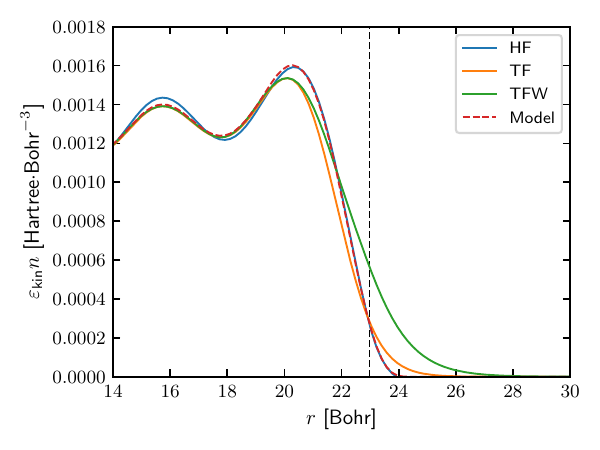}
\caption{\label{fig:eps_kin_comparison} Comparison of the local model of the total kinetic energy density $\varepsilon_\text{kin}$ for parameters $D = 2.89$, $F = 1.28\times10^{-4}$, $G = 1.68$, and $\omega = 1.60$ with TF and TFW for $n_\text{I} = 10^{-2}$ and $N = 508$. Vertical dashed line represents the boundary $R$.}
\end{figure}

A comparison of the model for the total kinetic energy density for $n_\text{I} = 10^{-2}$ with the Thomas--Fermi (TF) and the gradient-improved Thomas--Fermi--von Weisz\"acker (TFW) functions is shown in Fig. \ref{fig:eps_kin_comparison}.

\section{Conclusion}
We have presented a detailed Hartree--Fock analysis of the ground-state electronic structure of the finite spherically symmetric jellium model. The primary objectives of this study were to determine the correct ordering of the energy-level sequence, accurately describe the various contributions to the total energy, and precisely quantify the effect of confinement on the local models for exchange and kinetic energy potentials.
\par
Our results show that for a general positive background density $n_\text{I}$, the sequence of energy levels for a many-electron system follows a pattern that maximizes the orbital quantum number $l$ for each principal quantum number $n$. This is in agreement with the one-electron system, where the hydrogenic degeneracy is lifted and higher orbital-momentum waves become energetically preferable once the positive nuclear charge is spread over a finite volume.
\par
By comparing the exact behavior of the exchange energy density with the naive estimates, given by setting $n = n_\text{I}$, we see that the naturally occurring oscillations of the electron density are essential to correctly predict the increase in the effect of the local exchange potential inside confined systems. The analysis of the surface region reveals that the exchange potential is even stronger there and cannot be well described by a power-law function. The kinetic energy density remains well characterized by the infinite jellium analog but differs significantly in the surface area. Therefore, new local models of the exchange and kinetic energy densities for a given system size and background charge density were presented. These depend solely on the electron charge density, yet provide a more accurate description for the studied systems than some gradient-improved models. Moreover, the presented models are immediately applicable to quantum hydrodynamic calculations of large spherical nanoparticles made from various materials, although certain important phenomena, such as correlation or spin-orbit coupling, still need to be addressed.
\par
This work establishes the foundation for an accurate description of spherically symmetric confined electronic systems. The proposed exchange and kinetic energy functionals offer potential improvements for classical or orbital-free DFT, quantum hydrodynamics, and semi-classical models.

\section{Methods}
\subsection{Discretization of the Wave Functions and the Hamiltonian Operator}
To determine the energy levels of the jellium model and the corresponding radial wave functions, we employ the \ti{Matrix Method} \cite{gomez2018matrix}. This numerical approach is based on discretizing the radial Schrödinger equation within a finite interval $(0, r_{\text{max}}]$, assuming that the wave functions vanish at $r_{\text{max}}$.  The interval is divided into $M$ subintervals of uniform length $\Delta r$. The radial function is discretized into a vector of length $M$, where the entries represent the function values at the grid points $r_k = k\Delta r$. For notational convenience, we denote the discrete representation of the function by the same symbol, that is, $u_{nl}[k] \equiv u_{nl}(r_k)$.
\par
The radial one-electron Hamiltonian is then formulated as a matrix $\mathbb{H} = \mathbb{T} + \mathbb{V}$. The eigenenergies of the system are obtained as the eigenvalues of $\mathbb{H}$, with the corresponding eigenvectors representing the wave functions. The potential term of the Hamiltonian is given by the diagonal matrix:
\begin{align}
    \mathbb{V} = \text{diag}\big(V_{\text{eff}}(r_1), V_{\text{eff}}(r_2), ... , V_{\text{eff}}(r_{\text{max}})\big),
\end{align}
where $V_{\text{eff}}$ is the effective potential defined as:
\begin{align}
    V_{\text{eff}}(r) = \frac{l(l+1)}{2r^2} + V(r),
\end{align}
with $V$ given in Eq.\ (\ref{eq:potential_R}) and $l$ denoting the orbital quantum number. The kinetic energy operator is approximated using the central second difference scheme, resulting in a tridiagonal matrix:
\begin{align}
    \mathbb{T} = \frac{1}{2(\Delta r)^2} \begin{pmatrix}
        2 & -1 & \dots & 0 \\
        -1 & 2 & \dots & 0 \\
        \vdots & \vdots & \ddots & \vdots \\
        0 & 0 & \dots & 2
    \end{pmatrix}.
\end{align}
This formulation reduces the solution of the radial Schrödinger equation to a standard matrix eigenvalue problem which can be solved efficiently using linear algebra routines.
\par
In the many-electron system, the electron-electron interaction is added to the hamiltonian matrix. This is achieved by discretizing the electron-electron interaction operators $\hat{V}_{\text{H}}$ and $\hat{V}_\text{x}$ defined by Eqs. (\ref{eq:Hartree_potential_def}) and (\ref{eq:radial_exchange_op_def}). Since $\hat{V}_{\text{H}}$ acts on the wave function $u^s_{nl}$ simply by multiplying it by a function $V_{\text{H}}$, it can be represented by a diagonal matrix:
\begin{align}
    \mathbb{V}_{\text{H}} = \text{diag}\big(V_{\text{H}}(r_1), V_{\text{H}}(r_2), ... , V_{\text{H}}(r_{\text{max}})\big).
\end{align}
In contrast, the exchange potential couples different wave functions at different positions, requiring a full non-trivial matrix representation. To simplify its notation, we partition it into smaller submatrices. The exchange matrix acting on $u^s_{nl}$ can then be expressed as:
\begin{align}
    \mathbb{V}_\text{x}^{ls} = -\sum_{(n',\,l')}N^s_{n'l'}\sum_{L=|l-l'|}^{l + l'}\begin{pmatrix}
    l' & L & l \\
    0 & 0 & 0
    \end{pmatrix}^2\mathbb{V}_{n'l's}^L,
\end{align}
where the elements of the submatrices $\mathbb{V}_{n'l's}^L$ are defined as:
\begin{align}
    \mathbb{V}_{n'l's}^L[k,q] = u^{s}_{n'l'}[k]\frac{\min(k,q)^L}{\max(k,q)^{L+1}}u^{s}_{n'l'}[q],
\end{align}
for $k, q \in \{1, 2, ..., M\}$. This approach once again reduces the solution of the Schrödinger equation to a diagonalization problem, now involving the Hamiltonian matrix $\mathbb{H}^{ls} = \mathbb{T} + \mathbb{V} + \mathbb{V}_{\text{H}} + \mathbb{V}^{ls}_\text{x}$ for each unique one-electron radial wave function $u^s_{nl}$ in its vector representation.

\subsection{Implementation of the Hartree--Fock Self-Consistent Cycle}
The solution for each $N$-electron system consists of three main steps: choosing the appropriate electron configuration by assigning a unique set of quantum numbers to each electron, constructing the initial wave functions, and iteratively solving the Hartree--Fock equations until convergence is reached.
\par
We perform the calculations in ascending order of $N$, starting with the directly computable opposite-spin two-electron solution. To avoid any spin contamination, only spin-symmetrical systems are considered. When the uppermost shell becomes fully occupied, a new energy level is chosen from the set of unoccupied states, determined by the combinations $(n, l)$, where for each $l \in \{0,...,l_{\text{max}}+1\}$ the lowest unused $n$ is taken. Here, $l_{\text{max}}$ denotes the highest orbital quantum numbers used so far.
\par
The initial wave functions are selected at each increment of $N$ as follows: If no new energy level is introduced (i.e., the added electron pair remains in the same shell), the solution of the $(N - 2)$-electron system is employed as the starting guess. If a higher energy level is required, the initial wave function is chosen as the two-electron solution of the jellium potential with the current radius $R$, but the total positive charge reduced to $Q = 2$.
\par
After each iteration $t$, the wave functions are updated using a weighted sum:
\begin{align}
    u^{s,\,t}_{nl} = \lambda_t \tilde{u}^{s,\,t}_{nl} + (1 - \lambda_t)u^{s,\,t-1}_{nl},
\end{align}
where $\tilde{u}^{s,\,t}_{nl}$ is the solution of the Hartree--Fock equations and $\lambda_t$ is the learning rate. To reduce oscillations and stabilize the computational process, $\lambda_t$ was tuned to the value $0.25$. Although the newly computed wave functions $\tilde{u}^{s,\,t}_{nl}$ are, in principle, orthogonal, numerical artifacts lead to small deviations from the exact orthogonality. To mitigate this issue and restore an orthonormal basis, we apply the Löwdin symmetric orthogonalization procedure \cite{lowdin1950non} to the updated functions $u^{s,\,t}_{nl}$.
\par
Calculations were carried out for systems with the first fully filled orbitals exceeding $N = 500$. This yielded distinct $N_{\text{max}}$ for the different positive background charge densities used, ranging from $508$ to $520$ electrons. In order to correctly capture the effective wave function range and simultaneously maximize computational efficiency, we set the parameter $r_{\text{max}}$ ⁡as:
\begin{align}
    r_{\text{max}} = \min\bigg(\frac{20-4\log n_\text{I}}{2^{1/3}r_s}R, (8+2\log n_\text{I})\,500^{1/3}r_s\bigg).
\end{align}
The interval $(0,r_{\text{max}}⁡]$ was discretized into $500$ grid points.
\par
Convergence was monitored using three distinct criteria: differences in one-electron energies, wave functions, and electron densities. The conditions for convergence are:
\begin{align}
    |\epsilon^t_{nl} - \epsilon^{t-1}_{nl}| < \varepsilon, \\
    1 - |\langle u^{s,\,t}_{nl}|u^{s,\,t-1}_{nl}\rangle| < \delta, \\
    \Big|\int_0^{+\infty}\big(u^{s,\,t}_{nl}(r)^2 - u^{s,\,t-1}_{nl}(r)^2\big)\frac{1}{r^2}dr\Big| < \gamma,
\end{align}
for every electron shell $(n, l)$, spin $s$, and iteration $t$, where the parameters $\varepsilon = 5 \times 10^{-4}$, $\delta = 1 \times 10^{-4}$, and $\gamma = 5 \times 10^{-4}$ were chosen to balance precision with computational efficiency. To avoid excessively long computations, we imposed a maximum of 50 iterations.
\par
All computed wave functions and energy eigenvalues are available at \cite{piro_2025_17857264}.

\section*{Acknowledgment}
This work was financially supported by the FerrMion project of the Czech Ministry of Education, co-funded by the EU, Project No.\ CZ.02.01.01/00/22\_008/0004591; by the Czech Science Foundation (GAČR), Project No.\ 21-05259S; and by the Student Grant Competition of the Czech Technical University in Prague (SGS), Project No.\ SGS24/146/OHK4/3T/14.

\bibliography{main}% Produces the bibliography via BibTeX.

\end{document}

\preprint{AIP/123-QED}

\title[Supplementary Material]{Supplementary Material:\\A Hartree-Fock Analysis of the Finite Jellium Model}% Force line breaks with \\

\author{Michael Píro}
\author{Jaroslav Hamrle} 
\affiliation{
Department of Solid State Engineering, 
Faculty of Nuclear Sciences and Physical Engineering, Czech Technical University, Trojanova 13, Prague, Czech Republic
}

\date{\today}% It is always \today, today,
             %  but any date may be explicitly specified

\maketitle

\renewcommand{\thesection}{S\arabic{section}}
\setcounter{section}{0}
\renewcommand{\theequation}{S\arabic{equation}}
\setcounter{equation}{0}
\renewcommand{\thefigure}{S\arabic{figure}}
\setcounter{figure}{0}

\section{Jellium potential derivation}
For a given charge density distribution $n_\text{I}(\tb{r})$, the electrostatic potential $V(\tb{r})$ is defined as: 
\begin{align}
    V(\tb{r})=-\int_{\mathbb{R}^3}\frac{n_\text{I}(\tb{r}')}{|\tb{r}'-\tb{r}|}d\tb{r}'.
\end{align}
We consider a spherically symmetric confined system with radius $R$, centered at the origin, with a uniform positive charge density:
\begin{align}
    n_\text{I}(\tb{r}') = 
    \begin{cases}
        n_\text{I} & |\tb{r}'| \leq R \\
        0 & |\tb{r}'| > R.
    \end{cases}
\end{align}
Due to spherical symmetry, the potential depends only on the radial distance $r=|\tb{r}|$. Without loss of generality, we orient the coordinate system so that $\tb{r}$ lies along the $z$-axis. Then for the distance between the vectors $\tb{r}'$, $\tb{r}$ we have:
\begin{align}
    |\tb{r}'-\tb{r}| = \sqrt{r'^2 + r^2 - 2r'r\cos\theta'},
\end{align}
where $r' =|\tb{r}'|$. The potential then becomes:
\begin{align}
    V(r)&=-\int_0^{2\pi}d\phi'\int_0^\pi\int_0^R \frac{n_\text{I}r'^2}{\sqrt{r'^2 + r^2 - 2r'r\cos\theta'}} dr'\sin\theta' d\theta' = \nonumber\\
        &=-\frac{2\pi n_\text{I}}{r}\int_0^Rr'\int_{(r'-r)^2}^{(r'+r)^2}\frac{1}{2\sqrt{t}}\,dtdr' = -\frac{2\pi n_\text{I}}{r}\int_0^Rr'(r'+r-|r'-r|) dr',
\end{align}
where substitution $t=r'^2 + r^2 - 2r'r\cos\theta$ was used in the second step. We now distinguish two cases:
\begin{enumerate}
    \item $r \leq R$:
        \begin{align}
            V(r) = -\frac{2\pi n_\text{I}}{r}\big(\int_0^r2r'^2 dr' + \int_r^R 2r'r dr'\big)=-\frac{2\pi n_\text{I}}{3}(3R^2-r^2),
        \end{align}
    \item $r > R$:
        \begin{align}
            V(r)=-\frac{2\pi n_\text{I}}{r}\int_0^R2r'^2 dr' = -\frac{4}{3}\pi R^3\frac{n_\text{I}}{r}.
        \end{align}
\end{enumerate}
Now we just use the fact that the positive background charge $Q = \frac{4}{3}\pi R^3n_\text{I}$ and arrive at:
\begin{align}
    V(r) = \begin{cases}
        \frac{Q}{2R^3}(r^2 - 3R^2) & r \leq R \\
        -\frac{Q}{r} & r > R. \label{eq:nano_potential}
    \end{cases}
\end{align}
\par
The electrostatic energy of the background $E_\text{I}$ can be calculated as:
\begin{align}
    E_\text{I} = -\frac{1}{2}\int_{\mathbb{R}^3}V(\tb{r})n_\text{I}(\tb{r})d\tb{r} = -2\pi n_\text{I}\int_0^RV(r)r^2dr.
\end{align}
Using Eq. \ref{eq:nano_potential} we arrive at the following relation:
\begin{align}
    E_\text{I} = \frac{3Q^2}{5R}.
\end{align}

\section{Radial electron-electron interaction potentials}
The total electron–electron interaction energy for a system of $N$ electrons $\{\psi_i\}^N_{i=1}$ is given by:
\begin{align}
    E_\text{e-e} = \frac{1}{2}\sum^N_{i=1}\langle\psi_i|\hat{V}_\text{H} + \hat{V}_\text{x}|\psi_i\rangle, \label{eq:interaction_energy}
\end{align}
where the Hartree and exchange potentials are defined as:
\begin{align}
    \hat{V}_\text{H}\psi_i(\tb{r}) &= \sum_{j = 1}^N\int_{\mathbb{R}^{3}}\psi_j^*(\tb{r}')V_2(\tb{r},\tb{r}')\psi_j(\tb{r}')d\tb{r}'\,\psi_i(\tb{r}), \label{eq:Hartree_potential_op} \\
    \hat{V}_\text{x}\psi_i(\tb{r}) &= -\sum_{j = 1}^N\delta_{s_i s_j}\int_{\mathbb{R}^{3}}\psi_j^*(\tb{r}')V_2(\tb{r},\tb{r}')\psi_i(\tb{r}')d\tb{r}'\,\psi_j(\tb{r}), \label{eq:exchange_potential_op}
\end{align}
with $s_i$ denoting the spin of the wave function $\psi_i$ and the two-electron potential $V_2$ defined as: 
\begin{align}
    V_2(\tb{r}, \tb{r}') = \frac{1}{|\tb{r} - \tb{r}'|}.
\end{align}
To exploit the spherical symmetry, we expand $V_2$ using the \ti{generating function of Legendre polynomials}:
\begin{align}
    V_2(\tb{r},\tb{r}') = \frac{1}{\sqrt{r^2 + r'^2 - 2rr'\cos\vartheta}} = \sum_{l=0}^{+\infty}P_l(\cos\vartheta)\frac{\min(r,r')^l}{\max(r,r')^{l+1}},
\end{align}
where $r = |\tb{r}|$, $r' = |\tb{r}'|$, and $\vartheta$ is the angle between $\tb{r}$ and $\tb{r}'$. Next, we rewrite the polynomials $P_l$ with the help of the \ti{addition theorem of spherical harmonics}:
\begin{align}
    P_l(\cos\vartheta) = \frac{4\pi}{2l+1}\sum_{m=-l}^l{Y_l^m}^*(\theta,\phi)Y_l^m(\theta',\phi'),
\end{align}
where $\theta$, $\phi$, and  $\theta'$, $\phi'$ are the angles of $\tb{r}$ and $\tb{r}'$ in their common spherical coordinates. From there, $V_2$ can be expressed as:
\begin{align}
    V_2(\tb{r},\tb{r}') = \sum_{l=0}^{+\infty}\sum_{m=-l}^l\frac{4\pi}{2l+1}{Y_l^m}^*(\theta,\phi)Y_l^m(\theta',\phi')\frac{\min(r,r')^l}{\max(r,r')^{l+1}}. \label{eq:two_electron_potential}
\end{align}
The one-electron wave functions are written in the separable form:
\begin{align}
    \psi^s_{nlm}(r, \theta, \phi) = \frac{1}{r}u^s_{nl}(r)Y_l^m(\theta, \phi). \label{eq:wave_function}
\end{align}
Inserting Eqs. (\ref{eq:two_electron_potential}), (\ref{eq:wave_function}), and converting the electron index summation to a sum over quantum numbers $(n', l', m', s')$, the Hartree potential becomes:
\begin{align}
     \hat{V}_\text{H}\psi^s_{nlm}(r,\theta,\phi) =&  \sum_{(n',\,l',\,m',\,s')}\sum^{+\infty}_{L=0}\sum^L_{M=-L}\frac{4\pi}{2L+1}\int_0^{+\infty}u^{s'}_{n'l'}(r')^2\frac{\min(r,r')^L}{\max(r,r')^{L+1}}dr'\cdot \nonumber \\
     \cdot& \int_{\Omega}{Y^{m'}_{l'}}^*(\theta',\phi')Y^M_L(\theta',\phi')Y^{m'}_{l'}(\theta',\phi')\sin\theta'd\theta'd\phi'\,{Y^M_L}^*(\theta,\phi)\psi^s_{nlm}(r,\theta,\phi).
\end{align}
The angular integral can be evaluated using the standard identity:
\begin{align}
    \int_\Omega& {Y_{l_1}^{m_1}}^*(\theta,\phi)Y_{l_2}^{m_2}(\theta,\phi)Y_{l_3}^{m_3}(\theta,\phi)\sin\theta d\theta d\phi = \nonumber \\
    &= (-1)^{m_1}\sqrt{\frac{(2l_1+1)(2l_2+1)(2l_3+1)}{4\pi}}\begin{pmatrix}
    l_1 & l_2 & l_3 \\
    0 & 0 & 0
    \end{pmatrix}\begin{pmatrix}
    l_1 & l_2 & l_3 \\
    -m_1 & m_2 & m_3
    \end{pmatrix}, \label{eq:3_spherical_int}
\end{align}
where $\begin{pmatrix}
    a & b & c \\
    d & e & f
    \end{pmatrix}$ denotes the \ti{Wigner $3$-j symbol}. We can write:
\begin{align}
     \hat{V}_\text{H}\psi^s_{nlm}(r,\theta,\phi) =&  \sum_{(n',\,l',\,m',\,s')}\sum^{+\infty}_{L=0}\sum^L_{M=-L}(-1)^{m'}(2l'+1)\sqrt{\frac{4\pi}{2L + 1}}\begin{pmatrix}
    l' & L & l' \\
    0 & 0 & 0
    \end{pmatrix}\cdot \nonumber \\
     \cdot& \begin{pmatrix}
    l' & L & l' \\
    -m' & M & m'
    \end{pmatrix}\int_0^{+\infty}u^{s'}_{n'l'}(r')^2\frac{\min(r,r')^L}{\max(r,r')^{L+1}}dr'{Y^M_L}^*(\theta,\phi)\psi^s_{nlm}(r,\theta,\phi).
\end{align}
Similarly, for the exchange potential, we have:
\begin{align}
    \hat{V}_\text{x}\psi^s_{nlm}(r,\theta,\phi) =&  -\sum_{(n',\,l',\,m')}\sum^{+\infty}_{L=0}\sum^L_{M=-L}(-1)^{m'}\sqrt{(2l'+1)(2l+1)}\sqrt{\frac{4\pi}{2L + 1}}\cdot \nonumber \\
     \cdot& \begin{pmatrix}
    l' & L & l \\
    0 & 0 & 0
    \end{pmatrix}\begin{pmatrix}
    l' & L & l \\
    -m' & M & m
    \end{pmatrix}\cdot \nonumber \\
    \cdot& \int_0^{+\infty}u^{s}_{nl}(r')u^{s}_{n'l'}(r')\frac{\min(r,r')^L}{\max(r,r')^{L+1}}dr'{Y^M_L}^*(\theta,\phi)\psi^s_{n'l'm'}(r,\theta,\phi).
\end{align}
\par
Since the angular part of the wave function does not change during the Hartree-Fock procedure and we want to express the electron-electron interaction potentials only as operators acting on the radial part of the wave function, we rewrite the expression for the interaction energy given by Eq. (\ref{eq:interaction_energy}) as:
\begin{align}
    E_\text{e-e} = \frac{1}{2}\sum_{(n,\,l,\,m,\,s)}\langle\psi^s_{nlm}|\hat{V}_\text{H} + \hat{V}_\text{x}|\psi^s_{nlm}\rangle = \frac{1}{2}\sum_{(n,\,l,\,s)}\langle u^s_{nl}|\sum_m\langle Y^m_l|\hat{V}_\text{H} + \hat{V}_\text{x}|Y^m_l\rangle|u^s_{nl}\rangle. \label{eq:el_el_energy}
\end{align}
Note that the $1/r$ prefactors in the wave functions cancel the Jacobian $r^2$, so we do not explicitly write them out and handle the final integration over $r$ as a simple 1D integration with no additional factors. Together with Eq. \ref{eq:3_spherical_int} we arrive at the expression:
\begin{align}
    \sum_m\langle Y^m_l|\hat{V}_\text{H}|Y^m_l\rangle|u^s_{nl}\rangle =& \sum_m\sum_{(n',\,l',\,m',\,s')}\sum^{+\infty}_{L=0}\sum^L_{M=-L}(-1)^{m + m'}(2l + 1)(2l' + 1)\cdot \nonumber \\
    \cdot& \begin{pmatrix}
    l & L & l \\
    0 & 0 & 0
    \end{pmatrix}\begin{pmatrix}
    l & L & l \\
    -m & M & m
    \end{pmatrix}\begin{pmatrix}
    l' & L & l' \\
    0 & 0 & 0
    \end{pmatrix}\begin{pmatrix}
    l' & L & l' \\
    -m' & M & m'
    \end{pmatrix}\cdot \nonumber \\
    \cdot& \int_0^{+\infty}u^{s'}_{n'l'}(r')^2\frac{\min(r,r')^L}{\max(r,r')^{L+1}}dr'\,u^s_{nl}(r).
\end{align}
Similarly, for the exchange potential we get:
\begin{align}
    \sum_m\langle Y^m_l|\hat{V}_\text{x}|Y^m_l\rangle|u^s_{nl}\rangle =& -\sum_m\sum_{(n',\,l',\,m')}\sum^{+\infty}_{L=0}\sum^L_{M=-L}(2l + 1)(2l' + 1)\cdot \nonumber \\
    \cdot& \begin{pmatrix}
    l' & L & l \\
    0 & 0 & 0
    \end{pmatrix}^2\begin{pmatrix}
    l' & L & l \\
    -m' & M & m
    \end{pmatrix}^2\cdot \nonumber \\
    \cdot& \int_0^{+\infty}u^s_{nl}(r')u^{s}_{n'l'}(r')\frac{\min(r,r')^L}{\max(r,r')^{L+1}}dr'\,u^{s}_{n'l'}(r).
\end{align}
\par
Several simplifications can be made now. First, the second row of the Wigner 3-j symbol must sum up to zero. Therefore, the only valid values for the number $M$ are zero for the Hartree potential and $m' - m$ for the exchange potential. Secondly, at this stage, we consider every electron shell $(n,l)$ to be fully filled, i.e. the summations over the magnetic quantum numbers $m$ and $m'$ use all values between $-l$ and $l$, and $-l'$ and $l'$, respectively. This allows us to use the following summation formulas for the Wigner 3-j symbols:
\begin{align}
    \sum^{l}_{m=-l}(-1)^{m}\begin{pmatrix}
    l & L & l \\
    -m & 0 & m
    \end{pmatrix} &= (-1)^{l}\sqrt{2l + 1}\,\delta_{L0},\\
    \sum^l_{m=-l}\sum^{l'}_{m'=-l'}\begin{pmatrix}
    l' & L & l \\
    -m' & m'-m & m
    \end{pmatrix}^2 &= 1
\end{align}
followed by the simple relation:
\begin{align}
    \begin{pmatrix}
    l & 0 & l \\
    0 & 0 & 0
    \end{pmatrix} = (-1)^{l}\frac{1}{\sqrt{2l + 1}}.
\end{align}
Finally, the numbers in the first row of the Wigner 3-j symbol must obey the triangular inequality, yielding limits on the summation over $L$. The resulting relations for the two potentials acting on a single electron from the shell $(n,l)$ are of the form:
\begin{align}
    \hat{V}^\text{(rad)}_\text{H}u^s_{nl}(r) = \sum_{(n',\,l',\,s')}(2l' + 1)\int_0^{+\infty}u^{s'}_{n'l'}(r')^2\frac{1}{\max(r,r')}dr'\,u^s_{nl}(r), \label{eq:Hartree_potential}
\end{align}
and
\begin{align}
    \hat{V}^\text{(rad)}_\text{x}u^s_{nl}(r) =& - \sum_{(n',\,l')}(2l' + 1)\sum^{l+l'}_{L=|l-l'|}\begin{pmatrix}
    l' & L & l \\
    0 & 0 & 0
    \end{pmatrix}^2\cdot\nonumber\\
    &\cdot\int_0^{+\infty}u^s_{nl}(r')u^{s}_{n'l'}(r')\frac{\min(r,r')^L}{\max(r,r')^{L+1}}dr'\,u^{s}_{n'l'}(r), \label{eq:exchange_potential}
\end{align}
where the radial one electron potentials are defined as:
\begin{align}
    \hat{V}^\text{(rad)}_\text{H} &= \frac{1}{2l + 1}\sum_m\langle Y^m_l|\hat{V}_\text{H}|Y^m_l\rangle, \\
    \hat{V}^\text{(rad)}_\text{x} &= \frac{1}{2l + 1}\sum_m\langle Y^m_l|\hat{V}_\text{x}|Y^m_l\rangle. \label{eq:ex_equiv}
\end{align}
In the main text, however, we use the same symbols for the radial potentials as for the full interaction potentials for notational simplicity.
\par
Lastly, if the shells are not fully filled, we consider the electrons equally distributed over all possible quantum numbers $m$. The result therefore stays the same with only the multiplicity changing from $(2l+1)$ to $N^s_{nl}$ denoting the actual number of electrons in the shell $(n,l)$ with spin $s$.

\section{Exchange and kinetic energy densities as functions of the electron density}
To find a formula for the exchange energy density as a function of the electron density, we start by defining the one-electron exchange energy field $\varepsilon_\text{x}$ from the total exchange energy:
\begin{align}
    E_\text{x} = \int_{\mathbb{R}^3}\varepsilon_\text{x}(\tb{r})n(\tb{r})d\tb{r}, \label{eq:eps_definition}
\end{align}
where the total electron density of the electrons with is defined in terms of the one-electron wave functions as:
\begin{align}
    n(\tb{r}) = \sum_{(n,\,l,\,m,\,s)}|\psi^s_{nlm}(\tb{r})|^2.
\end{align}
In the case of not fully filled orbitals, we consider the wave functions to be equally distributed over all possible quantum numbers $m$. Therefore, we can assume spherical symmetry for all cases, and hence the electron density depends only on the radial coordinate $r$:
\begin{align}
    n(r) \equiv n(r, \theta, \phi) = \sum_{(n,\,l,\,m,\,s)}\frac{1}{r^2}u^s_{nl}(r)^2|Y^m_l(\theta, \phi)|^2 = \sum_{(n,\,l,\,s)}N^s_{nl}\frac{u^s_{nl}(r)^2}{4\pi r^2}.
\end{align}
Using this, we can rewrite Eq. \ref{eq:eps_definition} in spherical coordinates as:
\begin{align}
    E_\text{x} = \sum_{(n,\,l,\,s)}N^s_{nl}\int_0^{+\infty}\varepsilon_\text{x}(r)u_{nl}^s(r)^2dr.
\end{align}
On the other hand, from the Eqs. (\ref{eq:el_el_energy}) and (\ref{eq:exchange_potential}) we can express the total exchange energy from our Hartree-Fock calculations as:
\begin{align}
    E_\text{x} =& \:\frac{1}{2}\sum_{(n,\,l,\,s)}N^s_{nl}\langle u_{nl}^s|\hat{V}_\text{x}^\text{(rad)}|u_{nl}^s\rangle = \nonumber \\
    =& -\frac{1}{2}\sum_{(n,\,l,\,s)}N^s_{nl}\sum_{(n',\,l')}N^s_{n'l'}\sum^{l+l'}_{L=|l-l'|}\begin{pmatrix}
    l' & L & l \\
    0 & 0 & 0
    \end{pmatrix}^2\cdot\nonumber\\
    &\cdot\int_0^{+\infty}\int_0^{+\infty}u^s_{nl}(r')u^{s}_{n'l'}(r')\frac{\min(r,r')^L}{\max(r,r')^{L+1}}dr'\,u^{s}_{nl}(r)u^{s}_{n'l'}(r)dr.
\end{align}
By comparing the relations for the total exchange energy, we arrive at the final expression for the one-electron exchange field:
\begin{align}
    \varepsilon_\text{x}(r) &= -\frac{1}{2}\Big[\sum_{(n,\,l,\,s)}N^s_{nl}\sum_{(n',\,l')}N^s_{n'l'}\sum_{L=|l-l'|}^{l + l'}\begin{pmatrix}
    l' & L & l \\
    0 & 0 & 0
    \end{pmatrix}^2 \cdot \nonumber\\
    &\cdot \int_0^{+\infty}u^s_{nl}(r')u^s_{n'l'}(r')\frac{\min(r, r')^L}{\max(r, r')^{L+1}}dr'\,u^s_{nl}(r)u^s_{n'l'}(r)\Big] \Big/ \sum_{(n,\,l,\,s)}N^s_{nl}u_{nl}^s(r)^2.
\end{align}
The next step is to find a good approximation of the exchange energy density as a function of the electron density, i.e. the function $\varepsilon_\text{x}(n^s)$ satisfying:
\begin{align}
    \varepsilon_\text{x}(n(r)) \approx \varepsilon_\text{x}(r).
\end{align}
The functional of the exchange energy then becomes:
\begin{align}
    E_\text{x}[n] = 4\pi\int_0^{+\infty}\varepsilon_\text{x}(n(r))n(r)r^2dr,
\end{align}
and we can define the exchange potential function $V_\text{x}(n)$ as:
\begin{align}
    V_\text{x}(n) = \frac{\delta E_\text{x}[n]}{\delta n} = \varepsilon_\text{x}(n) + n\frac{d\varepsilon_\text{x}(n)}{d n}.
\end{align}
\par
Similarly, we can define the one-electron kinetic energy density $\varepsilon_\text{kin}$ via the total kinetic energy $E_\text{kin}$:
\begin{align}
    E_\text{kin} = \int_{\mathbb{R}^3}\varepsilon_\text{kin}(\tb{r})n(\tb{r})d\tb{r} = \sum_{(n,\,l,\,s)}N^s_{nl}\int_0^{+\infty}\varepsilon_\text{kin}(r)u_{nl}^s(r)^2dr.
\end{align}
From the Hartree-Fock approximation, we can rewrite the total kinetic energy as follows:
\begin{align}
    E_\text{kin} &= \sum_{i=1}^N\langle\psi_i|\hat{T}|\psi_i\rangle = -\frac{1}{2}\sum_{(n,\,l,\,s)}N^s_{nl}\Big\langle u_{nl}^s\Big|\frac{\partial^2}{\partial r^2 } - \frac{l(l+1)}{r^2}\Big|u_{nl}^s\Big\rangle = \nonumber \\
    &=-\frac{1}{2}\sum_{(n,\,l,\,s)}N^s_{nl}\int_0^{+\infty}\Big(u_{nl}^s(r)\frac{\partial^2u_{nl}^s(r)}{\partial r^2 } - \frac{l(l+1)}{r^2}u_{nl}^s(r)^2\Big)dr.
\end{align}
Therefore, the kinetic density can be expressed as:
\begin{align}
    \varepsilon_\text{kin}(r) &= -\frac{1}{2}\Big[\sum_{(n,\,l,\,s)}N^s_{nl}\Big(u_{nl}^s(r)\frac{\partial^2u_{nl}^s(r)}{\partial r^2 } - \frac{l(l+1)}{r^2}u_{nl}^s(r)^2\Big)\Big]\Big/\sum_{(n,\,l,\,s)}N^s_{nl}u_{nl}^s(r)^2.
\end{align}